\documentclass{article}
\usepackage{spconf,amsmath,epsfig}
\usepackage{subfigure}
\usepackage{algorithm,algpseudocode}
\usepackage{multirow}
\usepackage{booktabs}
\usepackage{amsfonts}
\usepackage{enumitem}
\usepackage{appendix}
\usepackage{color}
\usepackage{hyperref}
\definecolor{darkred}{rgb}{0.7,0,0}
\definecolor{darkgreen}{rgb}{0,0.5,0}
\definecolor{darkblue}{rgb}{0,0.0,0.3}
\hypersetup{colorlinks=true,
        linkcolor=darkred,
        citecolor=darkgreen
        }

\begin{document}\sloppy

\def\x{{\mathbf x}}
\def\L{{\cal L}}

\title{Tiyuntsong: A Self-Play Reinforcement Learning Approach for ABR Video Streaming}
%
\name{Tianchi Huang, Xin Yao, Chenglei Wu, Rui-Xiao Zhang, Zhengyuan Pang, Lifeng Sun}
\address{Dept. of Computer Science and Technology, Tsinghua University, Beijing, China \\
\{htc17,yaox16,wucl18,zhangrx17,pangzy12\}@mails.tsinghua.edu.cn, sunlf@tsinghua.edu.cn}
\maketitle

\begin{abstract}
Existing reinforcement learning~(RL)-based adaptive bitrate~(ABR) approaches outperform the previous fixed control rules based methods by improving the Quality of Experience~(QoE) score, as the QoE metric can hardly provide clear guidance for optimization, finally resulting in the unexpected strategies. In this paper, we propose \emph{Tiyuntsong}, a self-play reinforcement learning approach with generative adversarial network~(GAN)-based method for ABR video streaming. Tiyuntsong learns strategies automatically by training two agents who are competing against each other. Note that the competition results are determined by a set of rules rather than a numerical QoE score that allows clearer optimization objectives. Meanwhile, we propose GAN Enhancement Module to extract hidden features from the past status for preserving the information without the limitations of sequence lengths. Using testbed experiments, we show that the utilization of GAN significantly improves the Tiyuntsong's performance. By comparing the performance of ABRs, we observe that Tiyuntsong also betters existing ABR algorithms in the underlying metrics.
\end{abstract}
\section{Introduction}
Recent years have witnessed a rapid growth of online video streaming applications and services~\cite{cisco}. To achieve smooth video playback under various network conditions, modern client-side video player adopts ABR algorithm to dynamically determine the bitrate of next video chunk to download to achieve high QoE scores including high video bitrate, low rebuffering, etc. Most of the approaches, such as throughput-based\cite{li2014probe}, buffer-based\cite{spiteri2016bola} and mixed schemes\cite{yin2015control,spiteri2018theory} employ fixed control rules which determine future video bitrates via carefully tuned strategies and thresholds. However, these approaches are often designed with strong assumptions of the real-world network conditions and heavily rely on the fine-tuned parameters, which results in sensitivities to network conditions and unexpected performances~(\S\ref{sec:ABR}). To address these problems, researchers~\cite{mao2017neural} have proposed to leverage reinforcement learning~(RL) methods to \emph{learn} an algorithm from scratch without any network presumptions. In particular, the state-of-the-art deep reinforcement learning~(DRL) scheme Pensieve~\cite{mao2017neural} outperforms existing ABRs in some settings~\cite{akhtar2018oboe}. These work tries to optimize a neural network towards a better QoE score, in which the fine-tuned parameters have significant impacts on the performance. However, through the experiment, we observe that RL-based method often obtains high QoE scores via some tricks due to the lack of guidance for optimization in QoE~(\S\ref{sec:trap}). As a result, despite its abilities to obtain higher numerical QoE scores, such training schemes may generate a strategy that doesn't meet the basic rules of the ABR algorithm.


Our key idea is to regard the reward as a rule instead of QoE metrics~(\S\ref{sec:selfplay}). The rule~(\S\ref{sec:rule}) is allowed to be constructed by any methods, such as a logistic and AI method, aiming to identify the better one from two candidates. Unlike previous work, the rule will highlight the priority of optimization to avoid occurring unexpected strategies. The novel idea requires a new RL framework to match it. Thus, we propose \emph{Tiyuntsong}\footnote{Tinyuntsong: Also named as \emph{Cloud Ascending Ladder}, a qinggong skill in the Chinese wuxia novel \emph{The Heaven Sword and Dragon Saber} by Jin Yong. The skill enables the user to travel at high speeds and leap to extreme heights by stepping one foot on the other one.}, a self-play RL method with GAN for ABR video streaming~(\S\ref{sec:tiyuntsong}). Tiyuntsong trains two agents simultaneously for generating a well-performed ABR algorithm under different network conditions. In details~(\S\ref{sec:training}), Tiyuntsong first uses two agents to provide the video streaming services on the same network condition and video content respectively. Next, it leverages the rule to determine who is the winner. Finally, it assigns the reward of each agent as \emph{\{win:1, lose:0\}} and updates the two agents' gradients. In brief, Tiyuntsong approaches a Nash equilibrium via the self-play method, whereas traditional RL methods diverge. 

Besides, we further present \emph{GAN Enhancement Module}~(\S\ref{sec:GEM}), a GAN-based method to extract hidden features from the past status that facilitate Tiyuntsong to store the information without the limitation of sequence length. During the training process, the model generates future hidden features based on current state and hidden features, and then estimates the probability of whether the hidden feature comes from the winning sample or not.

In the rest of our paper~(\S\ref{sec:evaluation}), first, we collect a large corpus of network traces from alternative public datasets for training and validating. Next, we leverage Elo-Rating~\cite{elo1978rating}, a classic rating-based system to compute the performance of Tiyuntsong via win rate~(\S\ref{sec:rule}). Finally, using a testbed experiment~(\S\ref{sec:expresults}), we first discuss Tiyuntsong's neural network architecture~(\S\ref{sec:tiyuntsongarch}). After that, we prove the importance of GAN Enhancement Module~(\S\ref{sec:tivsgan}). In all considered scenarios, Tiyuntsong betters existing ABR approaches in both win rate and underlying metrics of ABR including bitrate and rebuffering as well as smoothness~(\S\ref{sec:tivsothers}).

To sum up, our contributions are as follows:

$\triangleright$~We figure out the weakness of RL-based ABR algorithms and suggest a novel sight to redefine the reward metric for them: leverage logistic rules instead of QoE metrics.

$\triangleright$~To the best of our knowledge, we are the first to use self-play RL method to tackle the ABR video streaming problem. Results indicate that Tiyuntsong not only avoids deviating from the fundamental rule but also betters recent work.

$\triangleright$~Traditional fusion of RL and GAN method~\cite{doan2018gan} pays more attention to improving the efficiency of imitation rather than preserving the useful information as described in this paper. In brief, we propose a novel perspective for the application of GAN, which also yields a reliable and effective result.

\section{Related Work and Motivation}

\subsection{ABR Algorithms Overview}
\label{sec:ABR}
Client-based ABR algorithms are mainly categorized into four types~\cite{bentaleb2018survey}: throughput-based, buffer-based, mixed and RL-based. PANDA~\cite{li2014probe} predicts the future throughput for eliminating the ON-OFF steady issue. 
However, due to the current lack of throughput estimation method, these approaches still result in poor ABR performance. 
Thus, many approaches are designed to select the proper bitrates based on playback buffer size observed. e.g., 
BOLA~\cite{spiteri2016bola} turns the ABR problem into a utility maximization problem and solve it by using the Lyapunov function. 
Nevertheless, the buffer-based approach fails to tackle the long-term bandwidth fluctuation problem.
~Then, to tackle the challenge, mixed approaches, such as MPC~\cite{yin2015control} and DynamicDASH~\cite{spiteri2018theory}, is proposed to select bitrate for next chunk by adjusting its throughput discount factor based on past prediction errors and predicting its playback buffer size. Note that such model-based approaches require careful tuning, because they rely on parameters that are quite sensitive to network conditions, which results in the poor performance in unexpected network environments.
To address these issues, several attempts~\cite{mao2017neural} have been made to optimize ABR algorithms based on RL method due to the difficulty of tuning mixed approaches for handling different network conditions. 

\begin{figure}
  \centering
  \begin{minipage}{1.0\linewidth}
    \centering 
    \includegraphics[width=0.49\linewidth]{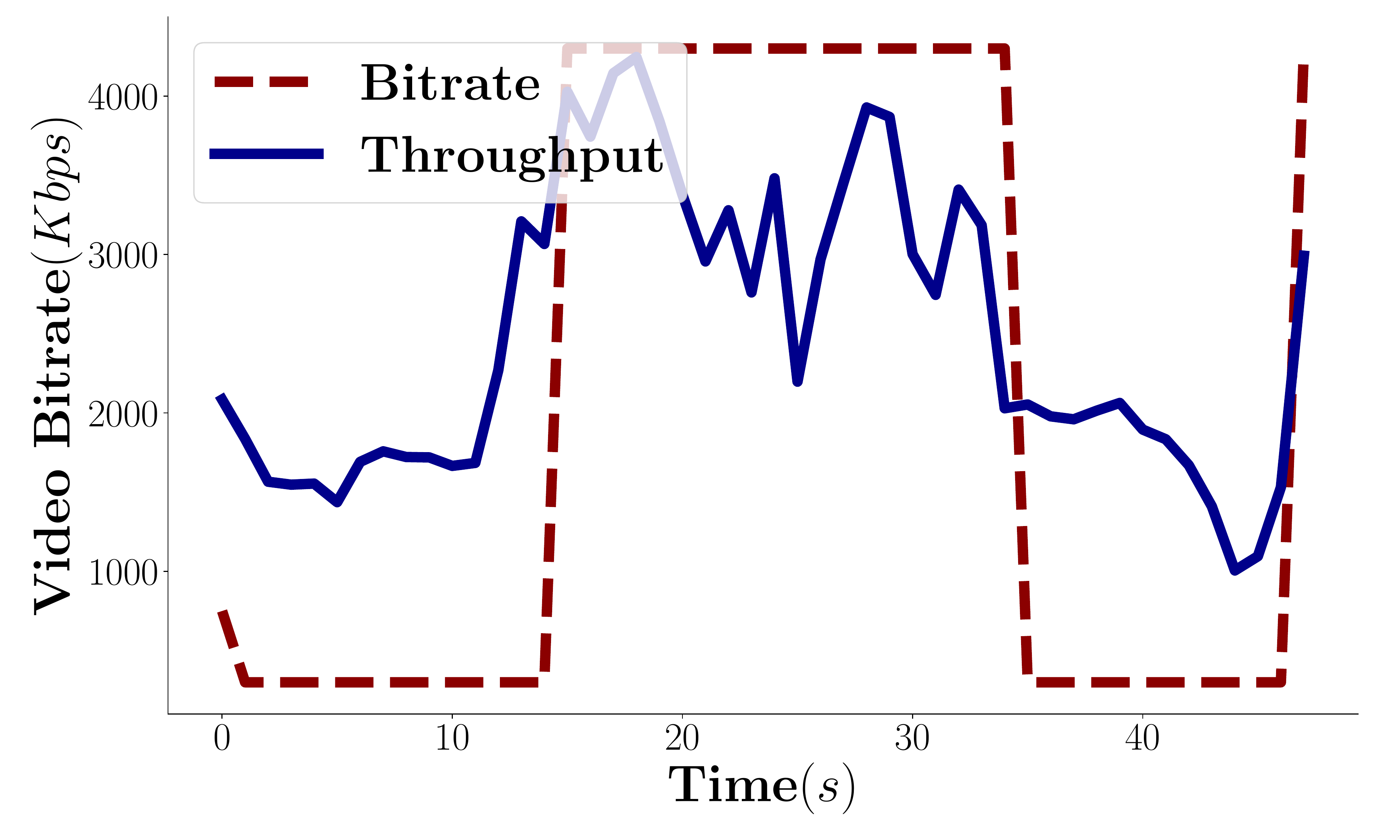}
    \includegraphics[width=0.49\linewidth]{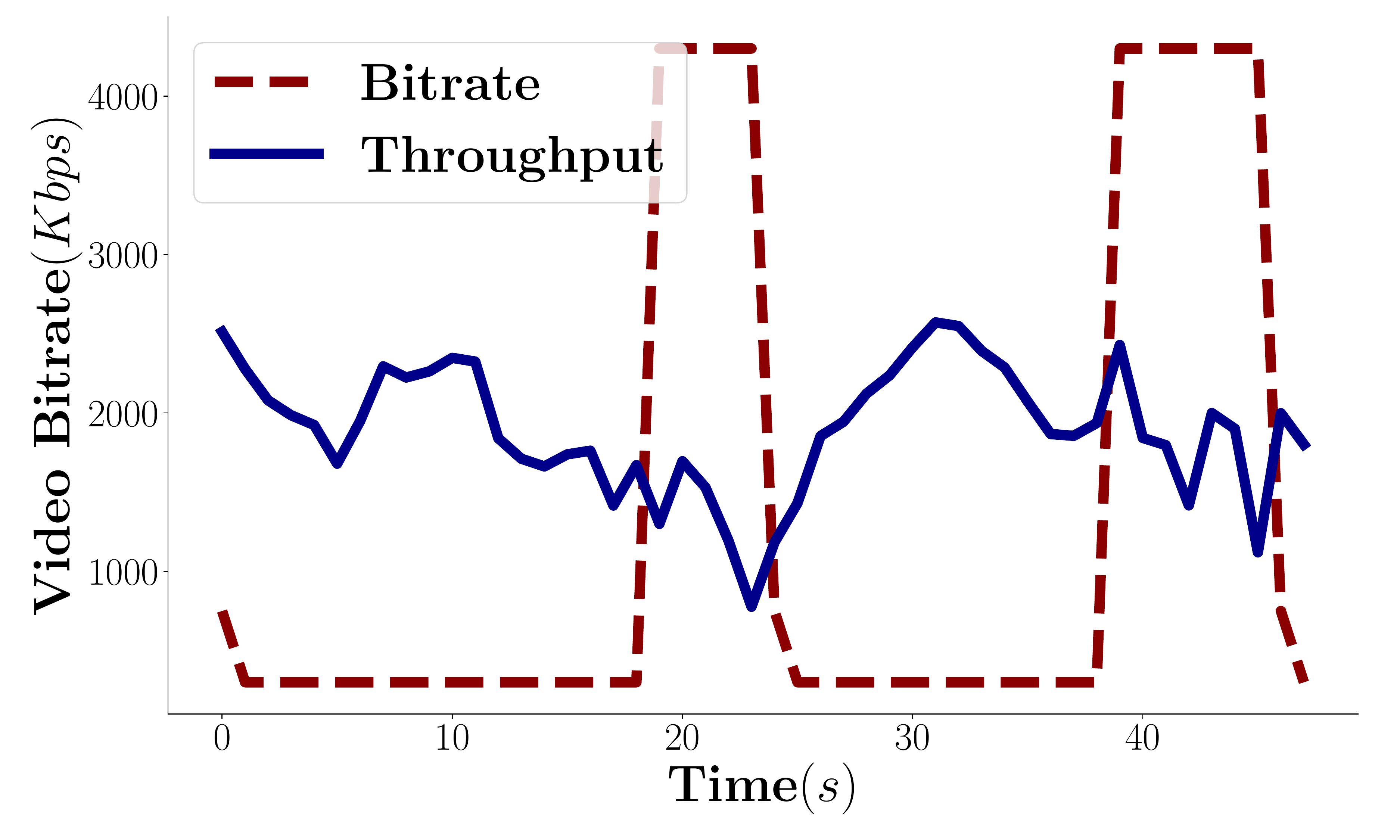}
      \caption{This group of figures shows the drawback of traditional RL-based ABR methods: After trained Pensieve with default parameter settings about seven days, we observed that:~1)~Its policy was simple and effective; 2)~It still maintained a high QoE score. However, it's clear that the proposed method deviates from the basic rules of ABR.}
  \label{fig:vmaf}
  \end{minipage}  
\end{figure}
\subsection{The Trap of Traditional RL-based Method}
\label{sec:trap}
However, traditional RL-based methods still have their drawbacks. Considering the ABR process as a Markov Decision Process~(MDP), in recent years, many schemes have been proposed to learn ABR algorithms via RL method~\cite{chiariotti2016online,mao2017neural}. Despite the outstanding performances that RL-based ABR algorithms achieve, these schemes suffer from a key limitation: They optimize their neural network via enhancing QoE scores. However, achieving a high QoE score doesn't necessarily equal to generate an exemplary algorithm. For example, the experimental results of state-of-the-art RL-based scheme Pensieve\footnote{\texttt{https://github.com/hongzimao/pensieve}} is illustrated in Figure~\ref{fig:vmaf}. 
While Pensieve converges, its policy can be generalized as 

\emph{Fetching lowest bitrates till the buffer has enough space and time to fetch highest bitrates}. 

That is because though RL methods will successfully train the model under the case which only needs to optimize one metric~(e.g., to play higher scores for Atari games), these schemes lack the abilities to tackle the problem affected by multiple factors directly. For example, in ABR problems, recent work~\cite{spiteri2016bola, yin2015control, mao2017neural} leverages reward functions ~(e.g., QoE metrics) with a weighted sum of several underlying metrics to take the place of the multi-factor optimization~(Several metrics must be optimized together). Hence, QoE driven RL-based approaches have the abilities to obtain a relative good numerical reward, but they may also provide users with unexpected and unstable viewing experience. This problem imposes critical challenges to RL-based ABR algorithm.

\subsection{Self-play Method}
\label{sec:selfplay}
AlphaZero~\cite{silver2017mastering}, a scientific breakthrough in AI, is trained solely based on self-play RL and periodically matched against several games. To tackle the traditional RL methods' problem, we thereby consider following AlphaZero to train the algorithm via self-play RL. However, static games with incomplete information (e.g., training ABR with two agents) are much dissimilar and more complicated than dynamic games with complete information such as Go. Thus, we meet new challenges:~\emph{How to design a proper model and how to define a suitable reward representation for ABR?}

\begin{figure}
    \centering
      \includegraphics[width=0.5\linewidth]{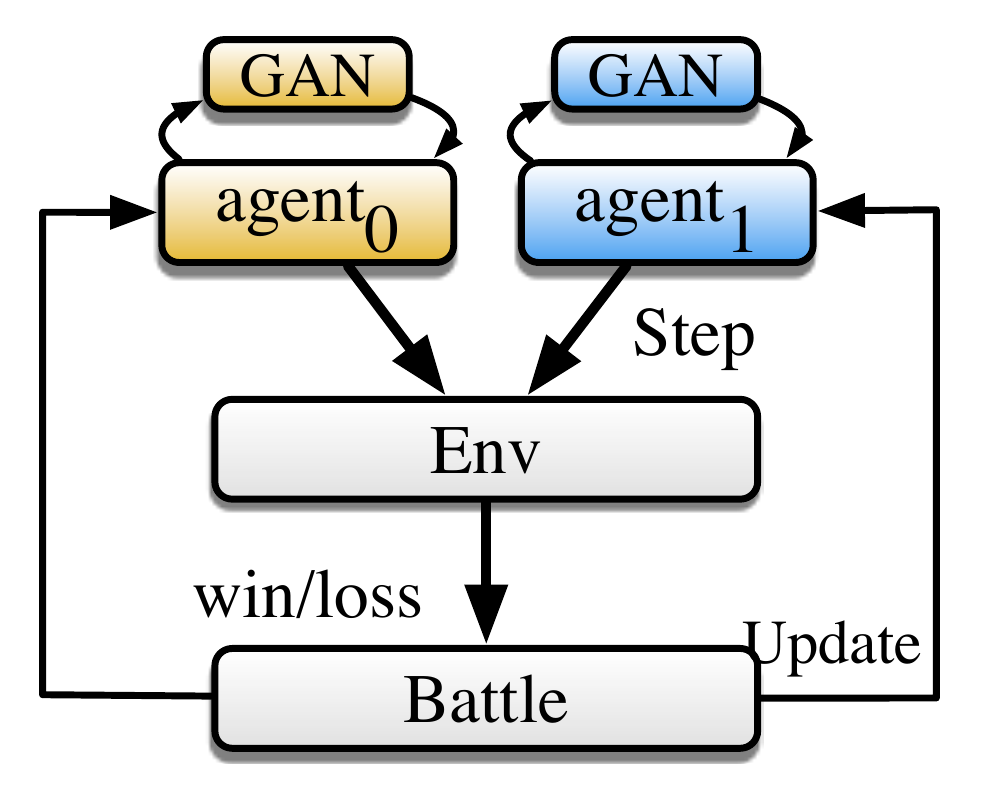}
    \caption{Tiyuntsong overview}
    \label{fig:overview} 
\end{figure}

\section{Tiyuntsong's Mechanism}
\label{sec:tiyuntsong}
In this section, we provide the design steps and the training methodology of Tiyuntsong~\footnote{https://github.com/thu-media/tiyuntsong}. As stated before~(\S\ref{sec:trap}), conventional RL is not a suitable scheme to solve the complex reward function problem in which its reward is computed as a linear combination of multiple factors. 
 We, therefore, suggest a novel sight to describe the reward function: Only to represent the reward as \emph{win} or \emph{loss} instead of an actual reward score. Following this sight, we propose Tiyuntsong, a self-play with RL method which learns an algorithm automatically based on the \emph{win or loss signal} only. As illustrated in Figure~\ref{fig:overview}, two agents compete for each other in the same environment and then update their network based on the competitive result.

\subsection{The Design of Agent}
We first initialize two agents $\mathbf{A_0}$ and $\mathbf{A_1}$. It is worth noting that Tiyuntsong's neural network architecture is quite different from the common RL's representation due to the distinctiveness of the ABR task.
The rest of the details are described as follows.

\noindent \textbf{State:} Tiyuntsong's learning agent takes the input state of time-slot $t$ $s_t = \{T,d,q,r,b, \mathbb{S}, h\}$ into neural network, where $T$ means the past throughput measured by a client for past $k$ sequence; $d$ represents the time for downloading past $k$ sequence; $q$ is the previous video bitrate selected of past $k$ sequence; $r$ is the remaining video playback time; $b$ is a buffer length used by the client; $\mathbb{S}$ is a vector that represents the video sizes of the next video chunk. The last one $h$ is a vector that reflects that extra features of the past, and it is generated by the GAN Enhancement Module~(See \S\ref{sec:GEM}).

\noindent \textbf{Action:} The action space is discrete. The output of the policy network is defined as a probability distribution: $f(s_t, a_t)$, meaning the probability of selection action $a_t$ being in state $s_t$. The action $a_t$ is an n-dims vector, which represents the candidate of video bitrate for the next chunk.


\noindent \textbf{Reward:}~Our reward is defined as a result: $r\in \{0,1\}$ judged by $\texttt{Rule}$. During the training process, we use $\texttt{Rule}$ to compute the win rate of two agents for each epoch, and in the result ``0'' means loss and ``1'' represents victory. We notice that $\texttt{Rule}$ can be represented as not only a human-made logistic algorithm but also a neural network model generalized by AI. The details of $\texttt{Rule}$ are described in \S\ref{sec:rule}. Based on the results calculated, we can estimate the win rate $w_i$ for each agent $\mathbf{A_i}$. We further utilize Elo Ratings to estimate the instant performance via win rate~(\S\ref{sec:elo}).

\subsection{GAN Enhancement Module}
\label{sec:GEM}
In recent work~\cite{sun2016cs2p,mao2017neural}, the lifetime of each bitrate decision has been modeled as an MDP~(Markov Decision Process), meaning that \emph{action} is only related to the status of the target rather than relying on the prior states. However, this assumption lacks evidence. \cite{sun2016cs2p} only illustrates that throughput factors can be efficiently captured by Hidden-Markov-Model~(HMM). Still, in \cite{mao2017neural}, the authors also consider different numbers of past throughout measurements to represent $state$. In general, previous work leverages a past $k$ steps status observed to reckon the status of the target in MDP, and the limitation of sequence length leads to missing crucial information of the past, such as the maximum and minimum values of the observed throughput.

\begin{figure}[ht]
    \centering
      \includegraphics[width=0.8\linewidth]{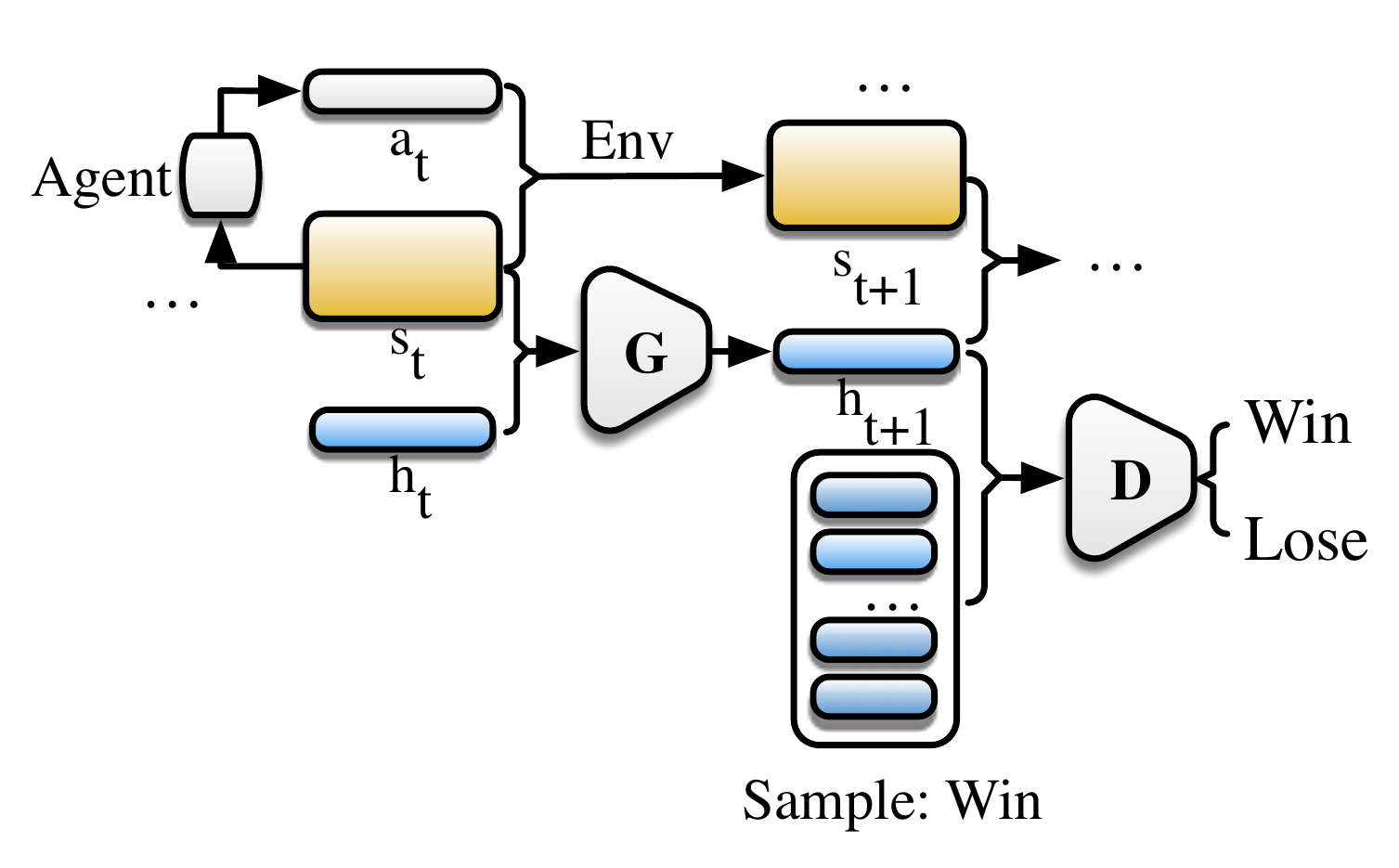}
    \caption{GAN Enhancement Module}
    \label{fig:gan} 
\end{figure}
Thus, we present \emph{GAN Enhancement Module} to automatically generate the hidden features from the past to break the limitation of sequence length. As illustrated in Figure~\ref{fig:gan}, the module consists of a generator~$\mathbf{G}$ and a discriminator~$\mathbf{D}$, where $\mathbf{G}$ is a function represented by several fully connected layers using \emph{leaky RelU} with parameters $\theta_g$, and $\mathbf{D}$ is also a function represented by multilayers with parameters $\theta_d$. For each step $t$, $\mathbf{G}$ is used for generating next hidden features $h_{t}$ according to the state $s_{t-1}$ and hidden feature $h_{t-1}$, and $\mathbf{D}$ outputs a single scalar $p_t \in [0,1)$ to estimate the probability that $h_{t}$ belongs to the historical winning samples. Inspired by LSGAN~\cite{mao2017least}, we first update $\mathbf{D}$ by descending its gradient according to $\mathbf{L}_d$~(Eq.~\ref{eq:dloss}). We then update $\mathbf{G}$ by descending its gradient via $\mathbf{L}_g$~(Eq.~\ref{eq:gloss}). Here $\mathbf{w}$ means the winning samples defined as $\mathbf{w}=\{h_0, h_1, \cdots,h_k\}$, where reward $r_k=1$.
\begin{small}
\begin{align}
    \begin{aligned}
    \begin{split}
        \min\limits_{\mathbf{D}}\mathbf{L}_d = &\frac{1}{2}E_{x\sim p_{w}(x)}[(\mathbf{D}(x) - 1)^{2}] + \\
            &\frac{1}{2}E_{t\sim p_{g}(t)}[(\mathbf{D}(\mathbf{G}(s_{t-1},h_{t-1})))^{2}]
    \end{split}
    \end{aligned}
    \label{eq:dloss}
\end{align}

\begin{align}
\min\limits_{\mathbf{G}}\mathbf{L}_g = \frac{1}{2}E_{t\sim p_{g}(t)}[(\mathbf{D}(\mathbf{G}(s_{t-1},h_{t-1})) - 1)^{2}]
\label{eq:gloss}
\end{align}
\end{small}

\subsection{Training Methodology}
\label{sec:training}
We now start to discuss how to train Tiyuntsong. In our work, we use the actor-critic method as the fundamental algorithm of Tiyuntsong. Each agent is composed of a policy network and a value network. 
The key thought of the policy gradient algorithm is to update the parameter in the direction of increasing the accumulated reward.
The gradient of the accumulated reward with respect to policy parameter $\theta$ can be written as

\begin{equation}
\nabla E_{\pi_{\theta}}[\sum_{t=0}^{\infty} \gamma^t r_t ]=E_{\pi_{\theta}}[\nabla_{\theta}\log_{\pi_{\theta}}(s,a)A^{\pi_{\theta}}(s,a)].  
\end{equation}

We can use~$E_\theta[\nabla_\theta log{\pi_\theta(s,a)}A^{\pi_\theta}(s,a)]$ as its unbiased form, where $A(s_t,a_t)$ is called the advantage of action $a_t$ in state $s_t$ which satisfies the following equality: $A(a_t, s_t)=Q(a_t,s_t)-V(s_t)$, where $V(s_t)$ represents the estimate of the value function of state $s_t$ and $Q(a_t, s_t)$ is the value of taking certain action at state $s_t$.  Next, we consider to use n-step Q-learning for optimizing the value network. The value network will be updated as

\begin{small}
\begin{equation}
\theta_v \gets \theta_v - \alpha_{v} \sum_t \nabla_{\theta_v} (\underbrace{r_t+\gamma V(s_{t+1}|\theta_v)}_{Q(a_t, s_t)}-V(s_t|\theta_v))^2. 
\label{eq:1}
\end{equation}
\end{small}

Here $V(s_t|\theta_v)$ is the estimation of $V(s_t)$; The direction of changing parameter $\theta_v$ is the negative gradient of it; $\alpha_v$ is the learning rate for the value network. We also add the entropy of policy in the object of policy network, which can effectively discourage the network to converge to sub-optimal policies. Thus, the update of $\theta$ will be written as

\begin{equation}
\begin{small}
\theta \gets \theta + \alpha_p \sum_t \nabla_\theta \log_{\pi_\theta}(s_t,a_t)A(s_t,a_t)+\beta \nabla_\theta H(\pi_\theta(\cdot|s_t)),
\end{small}
\label{eq:2}
\end{equation}

\noindent where $H(\cdot)$ is the entropy of the policy. After convergence, the value network will be abandoned, and we only use policy network to make decisions;  $\alpha_p$ is a learning rate function; $\beta$ is a hyper-parameter regarded as the weight of exploration. For each epoch $i(i>0)$, the parameters $\alpha_{p_i}$ and $\alpha_{v_i}$ can be computed by the equalization as follows:
\begin{small}
\begin{equation}
\alpha_{p_i}, \alpha_{v_i}=
\begin{cases}
(\alpha_{p_0}, \alpha_{v_0})(w_i\log w_i + 2.0) & \text{$w_i<0.5$}\\
-(\alpha_{p_0}, \alpha_{v_0})w_i\log w_i & \text{$w_i \geq 0.5$},
\end{cases}
\end{equation}
\end{small}

\noindent in which $w_i$ is the win rate for each training epoch $i$; $\alpha_{p_0}$ and $\alpha_{v_0}$ are initialized hyper-parameters which control the overall learning rate of policy network and value network. Dynamic learning rate can effectively avoid a considerable gap between the two agents. 


\section{Evaluation}
\label{sec:evaluation}
\subsection{Experimental Setup}

\textbf{Datasets}~We collect about 2,300 network traces from different public datasets for training and evaluating Tiyuntsong. The details of our network traces are composed of~Norway~\cite{riiser2013commute},~Synthetic Network Traces~\cite{mao2017neural},~Belgium~\cite{van2016http},~FCC~\cite{mao2017neural}, and Oboe~\cite{akhtar2018oboe}. 

\begin{table}[ht]
    \caption{The $\texttt{rule}$ Used In The Experiment}
    
    \begin{minipage}{1.0\linewidth}
        \subfigure[]{
        \centering
        \begin{tabular}{p{0.18\linewidth}|p{0.18\linewidth}|p{0.18\linewidth}|p{0.18\linewidth}}
            \toprule
            \textbf{\texttt{Rule}} & $b_0 > b_1$ & $b_0 = b_1$ & $b_0 < b_1$ \\
            \hline
            $r_0 > r_1$ & Table~1(b) & $\mathbf{1}$ & $\mathbf{1}$ \\
            $r_0 = r_1$ & $\mathbf{0}$ & Table~1(c) & $\mathbf{1}$ \\
            $r_0 < r_1$ & $\mathbf{0}$ & $\mathbf{0}$ & Table~1(b) \\
            \bottomrule
        \end{tabular}
        }
    \end{minipage}
    \begin{minipage}{0.60\linewidth}
        \subfigure[]{
            \centering
            \begin{tabular}{p{0.5\linewidth}|p{0.25\linewidth}}
                \toprule
                ${r_0}/{b_0} > {r_1}/{b_1}$ & $\mathbf{1}$ \\
                
                ${r_0}/{b_0} = {r_1}/{b_1}$ & $\mathbf{0}$ \\
                
                ${r_0}/{b_0} < {r_1}/{b_1}$ & Table~1(c) \\
                \bottomrule
            \end{tabular}
        }
    \end{minipage}
    \hfill
    \begin{minipage}{0.35\linewidth}
    
        \subfigure[]{
            \centering
            \begin{tabular}{p{0.4\linewidth}|p{0.2\linewidth}}
                \toprule
                $s_0 > s_1$ & $\mathbf{1}$ \\
                \hline
                $s_0 = s_1$ & $\mathbf{0}/\mathbf{1}$ \\
                $s_0 < s_1$ & $\mathbf{0}$\\
                \bottomrule
            \end{tabular}
        }
    \end{minipage}
    \label{alg:rule}
\end{table}

\label{sec:rule}
\noindent \textbf{The Design of \texttt{Rule}}~A good ABR algorithm mainly consists of three underlying metrics~\cite{spiteri2018theory}: 

$\triangleright$~\textbf{Bitrate:} To play the video at the highest sustainable quality, such as bitrate and video quality. 

$\triangleright$~\textbf{Rebuffering:} To avoid rebuffering events that occur due to the empty client buffer. 

$\triangleright$~\textbf{Smoothness:} Keep the bitrate in little change during the entire session. 

To this end, we implement an intuitive logistic rule\footnote{We repeat that \texttt{Rule} is allowed to design in any way, i.e., logistic or AI methods, etc. In this paper, due to the space limitations, we only give an intuitive rule for evaluating Tiyuntsong because too many statements about the rules will be badly miscast here.} for evaluation based on their priority~(See in Table~\ref{alg:rule}), in which $b_i$ denotes total bitrate, $r_i$ is total rebuffer time and $s_i$ represents total bitrate change for each agent $i\in\{0,1\}$.  


\noindent \textbf{Metrics}~
\label{sec:elo}
We leverage Elo Ratings~\cite{elo1978rating}, a traditional method for calculating the relative performance of players in zero-sum games, to evaluate Elo Ratings based on win rate. We first select several previously proposed approaches and test their performance respectively under the same environment. Next, we use \texttt{rule} to estimate their win rate. Finally, we compute the Elo rating for these approaches. In our work, these scores are defined as \emph{baselines}. For each epoch, the agent compares the result with baselines and then computes the Elo rating through the win rate. In this experiment, we set hyper-parameter $K=10$ and initialized rating $I=1000$ for the Elo system. More details are described in our repo..


\noindent \textbf{Testbed Setup}~We utilize Sabre~\cite{spiteri2018theory}, a state-of-the-art open-sourced simulation environment for ABR algorithms, to precisely emulate the ABR's process in an offline environment. Sabre is a Python tool that can quickly evaluate ABR algorithms in an emulated environment similar to real production players. For each step $t$, the agent uses the Sabre environment to simulate the entire session with given video descriptions and network traces.

\subsection{Experiments and Results}
\label{sec:expresults}




\begin{table}
    \centering
        \begin{small}
        \begin{tabular}{c|c|c}
            \toprule
            \textbf{Arch.} & \textbf{Elo} & \textbf{Timespan(it/s)}\\
            \hline
            FC & 1033 & \textbf{1.28} \\
            LSTM & 1057 &  0.77 \\
            2D-CNN & 1040 & 1.16 \\
            1D-CNN & \textbf{1094} & 1.04 \\
            Constrained & 977 & - \\
            Throughput-Rule & 1023 & - \\
            \bottomrule
        \end{tabular}
        \end{small}
        \caption{Comparing performance (Elo Ratings) of Tiyuntsong with different neural network architectures including Fully Connected, 2D-CNN, 1D-CNN and LSTM. Results are evaluated under same network traces and video description in 50 steps.}
        \label{fig:tytarch}
\end{table}

\subsubsection{Tiyuntsong with Different Architectures}
\label{sec:tiyuntsongarch}
In this experiment, we compare Tiyuntsong's network architecture to the following network architectures which collectively represent the architecture candidates: 

\begin{small}
\textbf{Fully Connected:} $\texttt{FC}_{64}^{1} \rightarrow \texttt{FC}_{128}^{2} \rightarrow \texttt{FC}_{64}^{3}$ 
    
\textbf{LSTM:} $\texttt{LSTM}_{64}^{1} \rightarrow \texttt{LSTM}_{64}^{2} \rightarrow \texttt{SELF-ATTENTION}_{64}^{1}$
    
\textbf{2D-CNN:} $\texttt{CONV2D}_{64}^{1} \rightarrow \texttt{MAXPOOL}_{2}^{1} \rightarrow \texttt{CONV2D}_{64}^{2} \rightarrow \texttt{MAXPOOL}_{2}^{2} \rightarrow \texttt{FC}_{64}^{1}$
    
\textbf{1D-CNN$^{*}$:} $\texttt{CONV1D}_{64}^{1\cdots6} \rightarrow \texttt{MERGE}^{1} \rightarrow \texttt{FC}_{64}^{1}$

\end{small}
We train and test under Sabre with the same network traces and video descriptions. In this experiment, we set $\gamma=0.99$, $\beta=0.02$, $step=50$ for only testing their performance instead of convergence. We report the result in Figure~\ref{fig:tytarch}, where 1D-CNN is the Tiyuntsong's network architecture. The obtained results indicate that 1D-CNN neural network architecture succeeds in improving the Elo Ratings, with improvements in average Elo Ratings of 37 - 61. We also observe that there is no obvious difference between these architectures regarding operational efficiency.

\subsubsection{Tiyuntsong vs. Tiyuntsong without GAN}
\label{sec:tivsgan}
\begin{figure}[ht]
    \centering
    \includegraphics[width=0.8\linewidth]{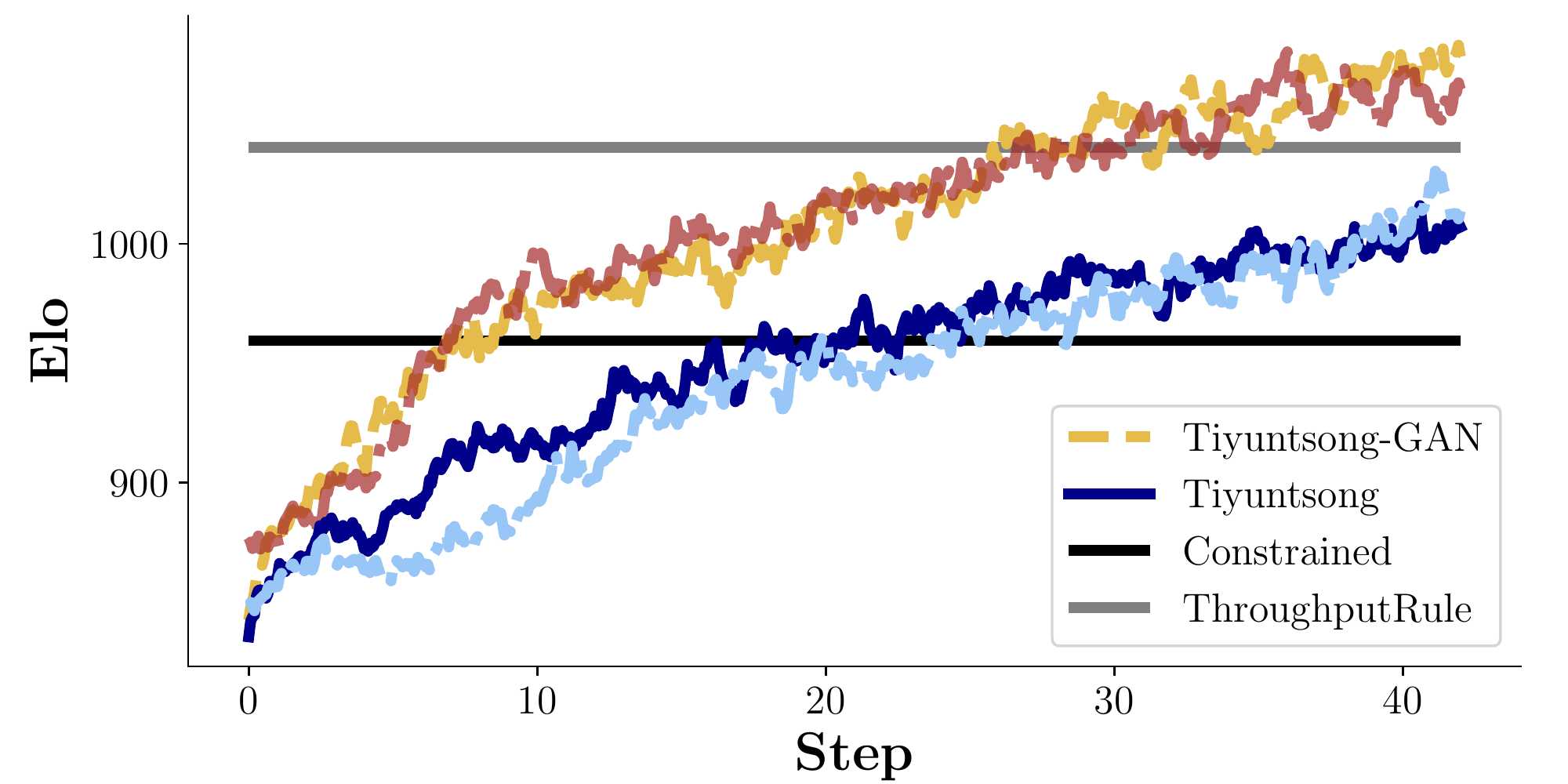}
    \caption{Comparing Tiyuntsong-GAN with Tiyuntsong on the same network traces. Results are evaluated in Elo Ratings based on previous approaches as baselines.}
    \label{fig:gan_vs_nogan} 
\end{figure}

In this part, we design an experiment to confirm whether the GAN Enhancement Module is effective or not. We set $step=50$ and compare Tiyuntsong-GAN with Tiyuntsong without using GAN Enhancement Module on the same network traces, and use two existing approaches: constrained and throughput rule as baselines. The experimental result is illustrated in Figure~\ref{fig:gan_vs_nogan}. As expected, we observe that Tiyuntsong-GAN outperforms Tiyuntsong with improvements in average Elo Ratings of 13.3\% after 50 steps.
\begin{figure}
    \centering
    \includegraphics[width=0.8\linewidth]{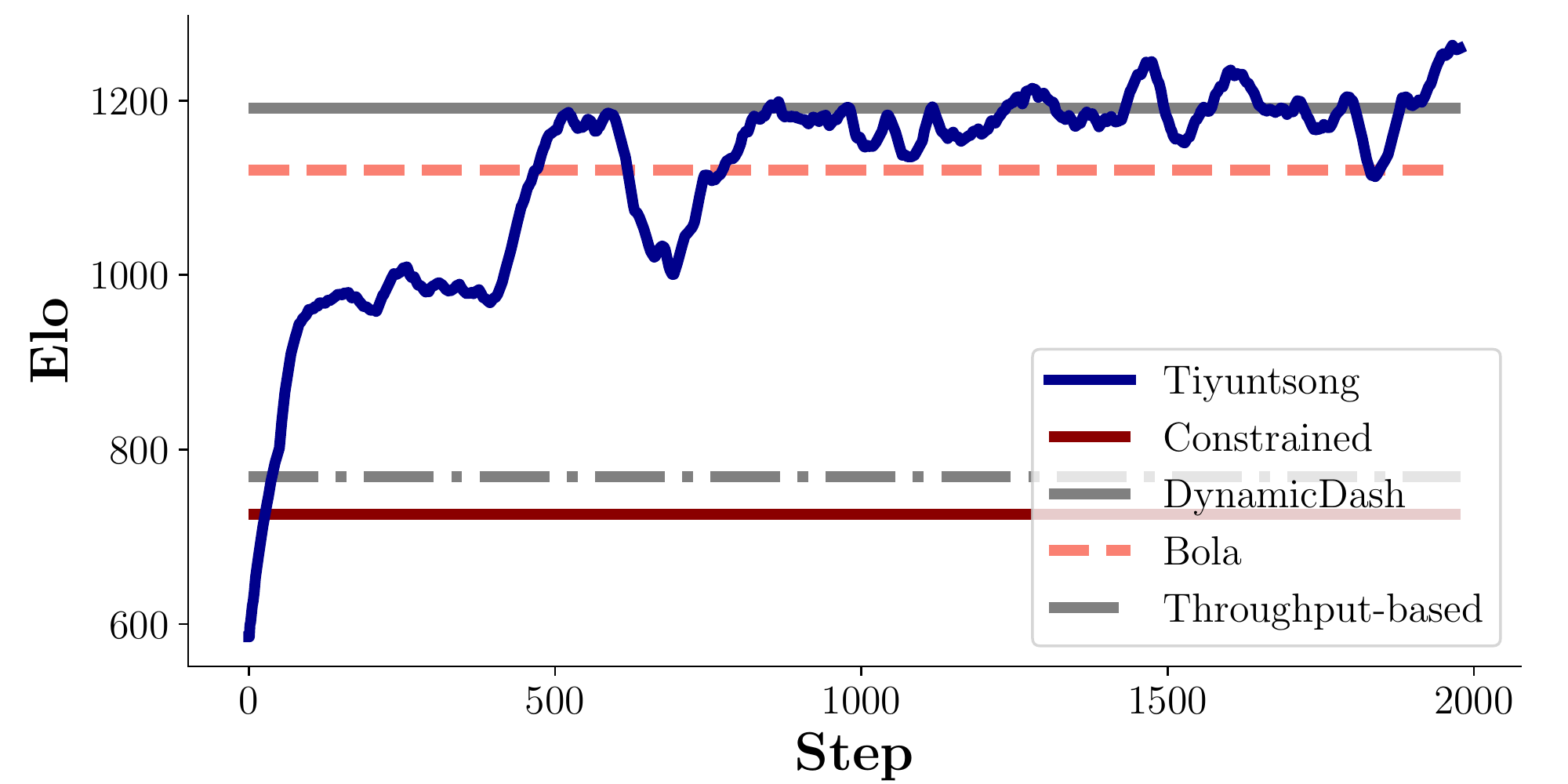}
    \caption{The curve of training Tiyuntsong for 2,000 steps. Elo-ratings are computed from $\texttt{Rule}$ between different ABR algorithms including Constrained, Throughput-Based, Bola and DynamicDASH.}
    \label{fig:tytvs} 
\end{figure}

\begin{figure}
    \centering
      \includegraphics[width=0.48\linewidth]{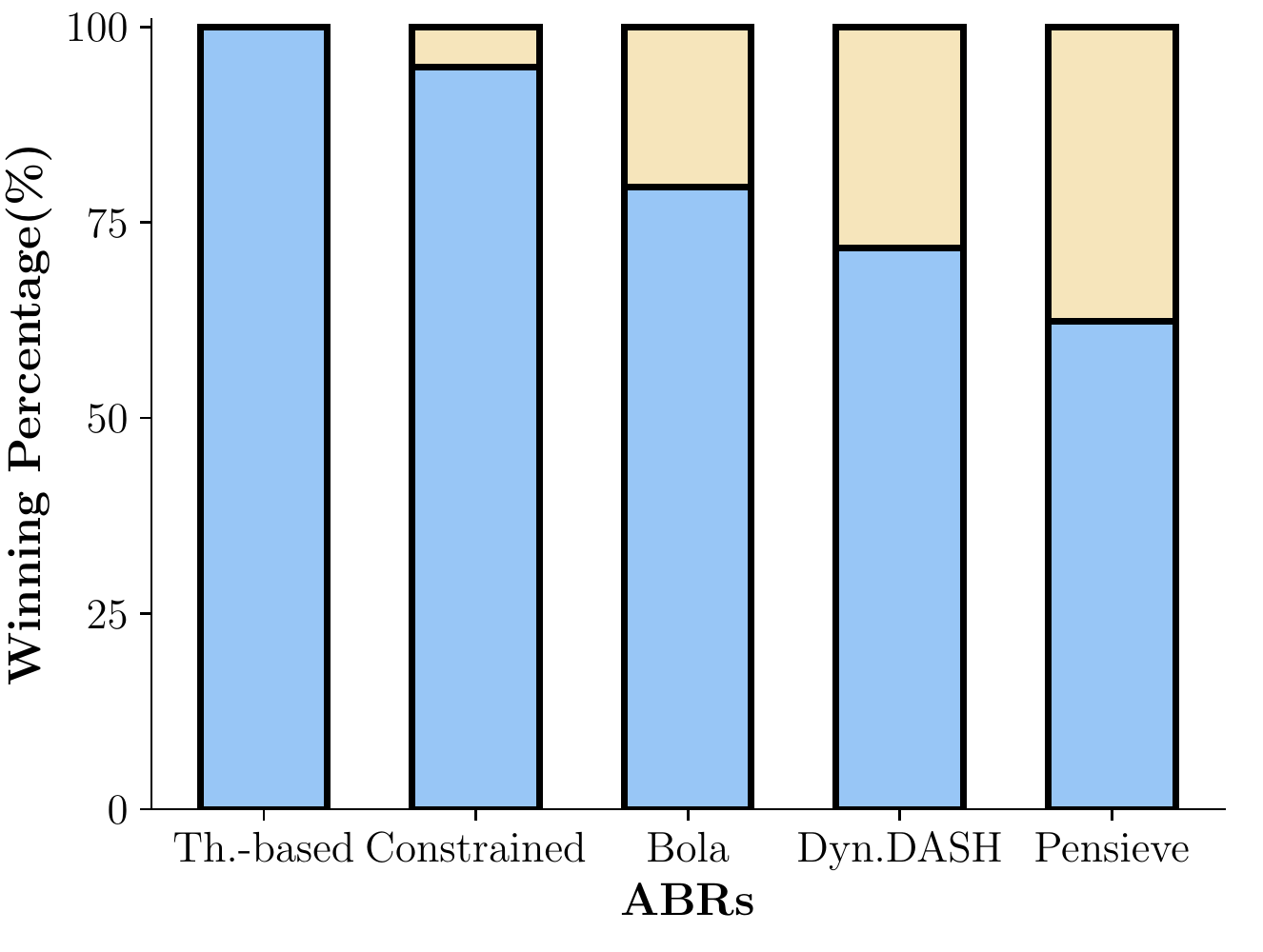}
      \includegraphics[width=0.48\linewidth]{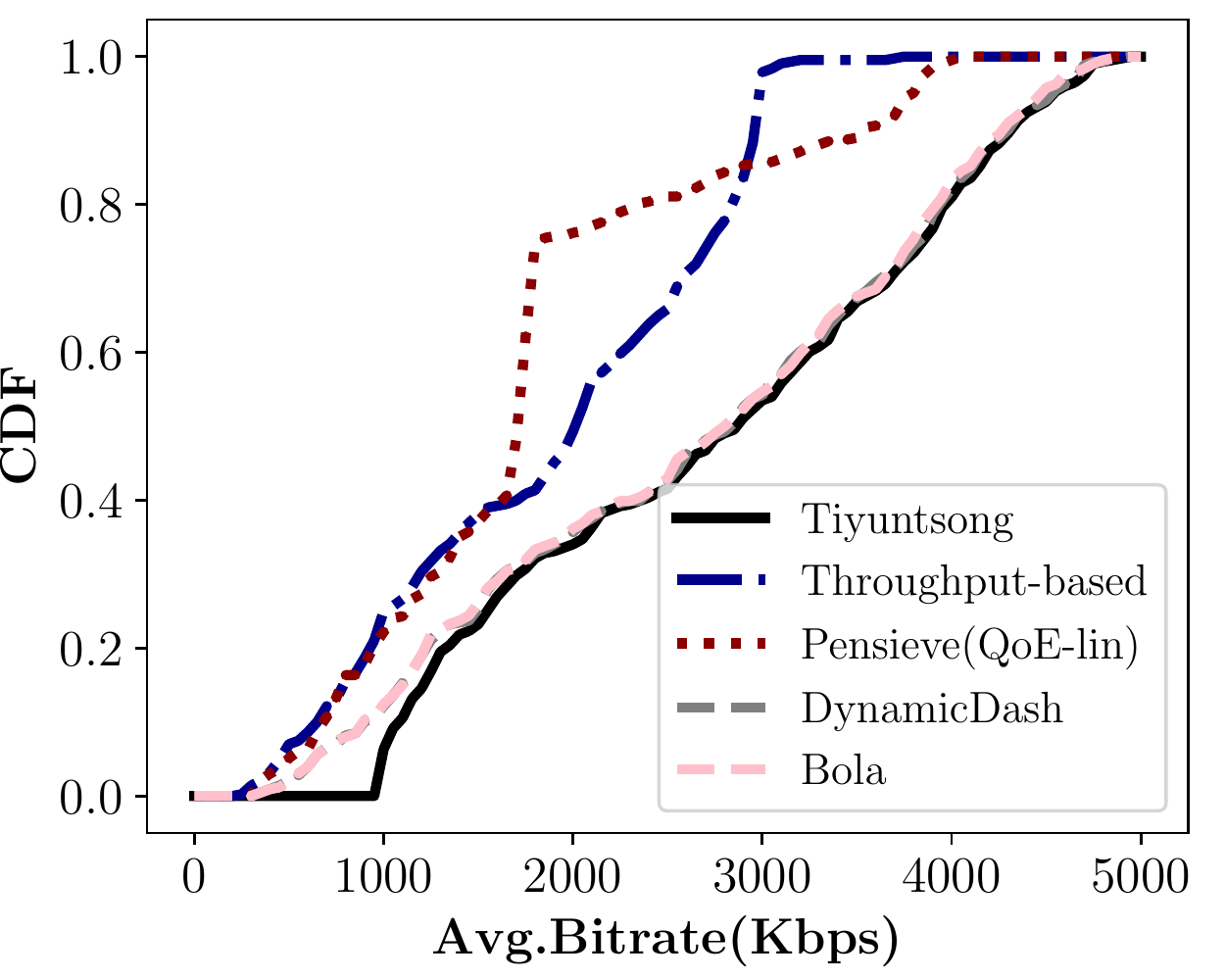}
      \includegraphics[width=0.48\linewidth]{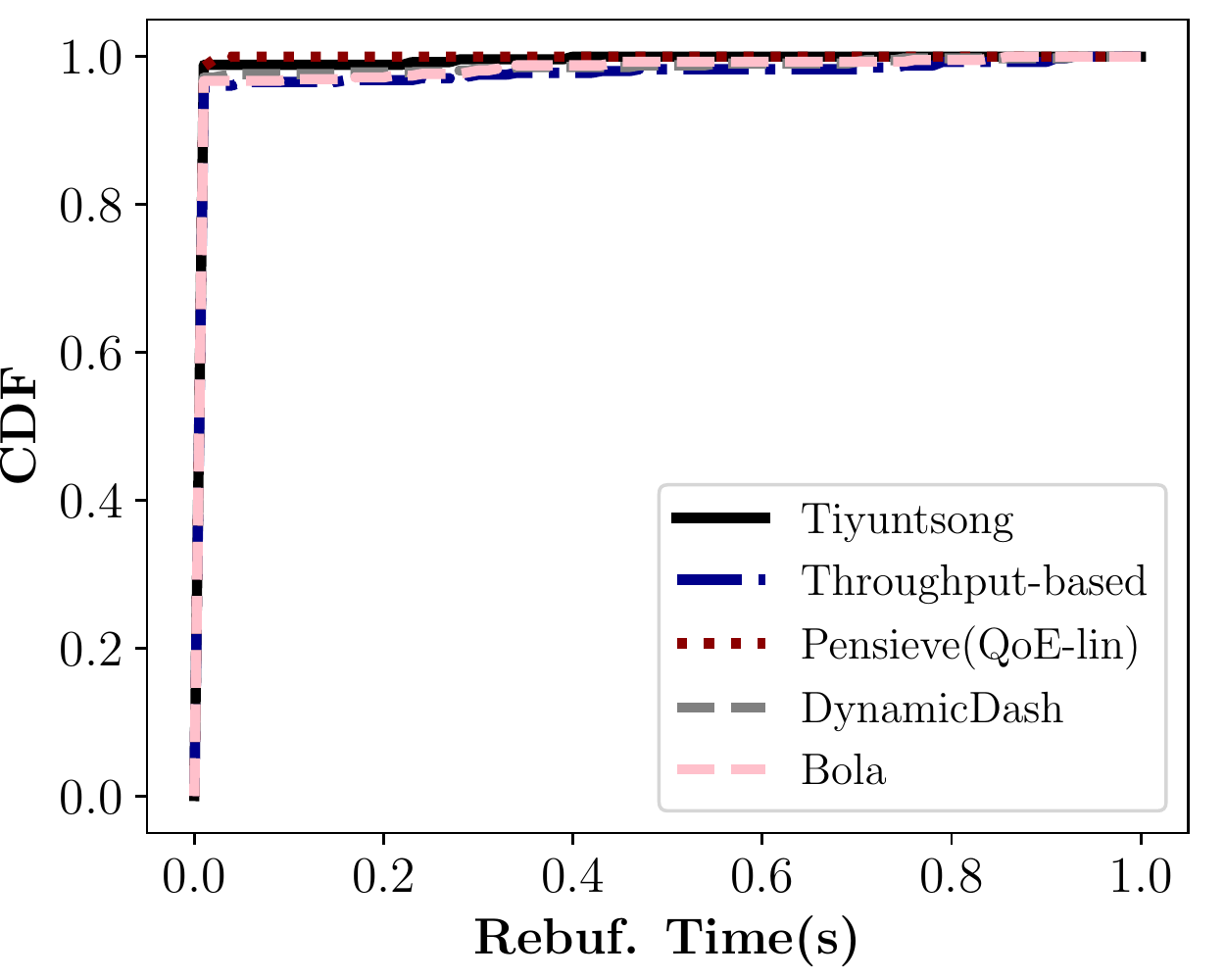}
      \includegraphics[width=0.48\linewidth]{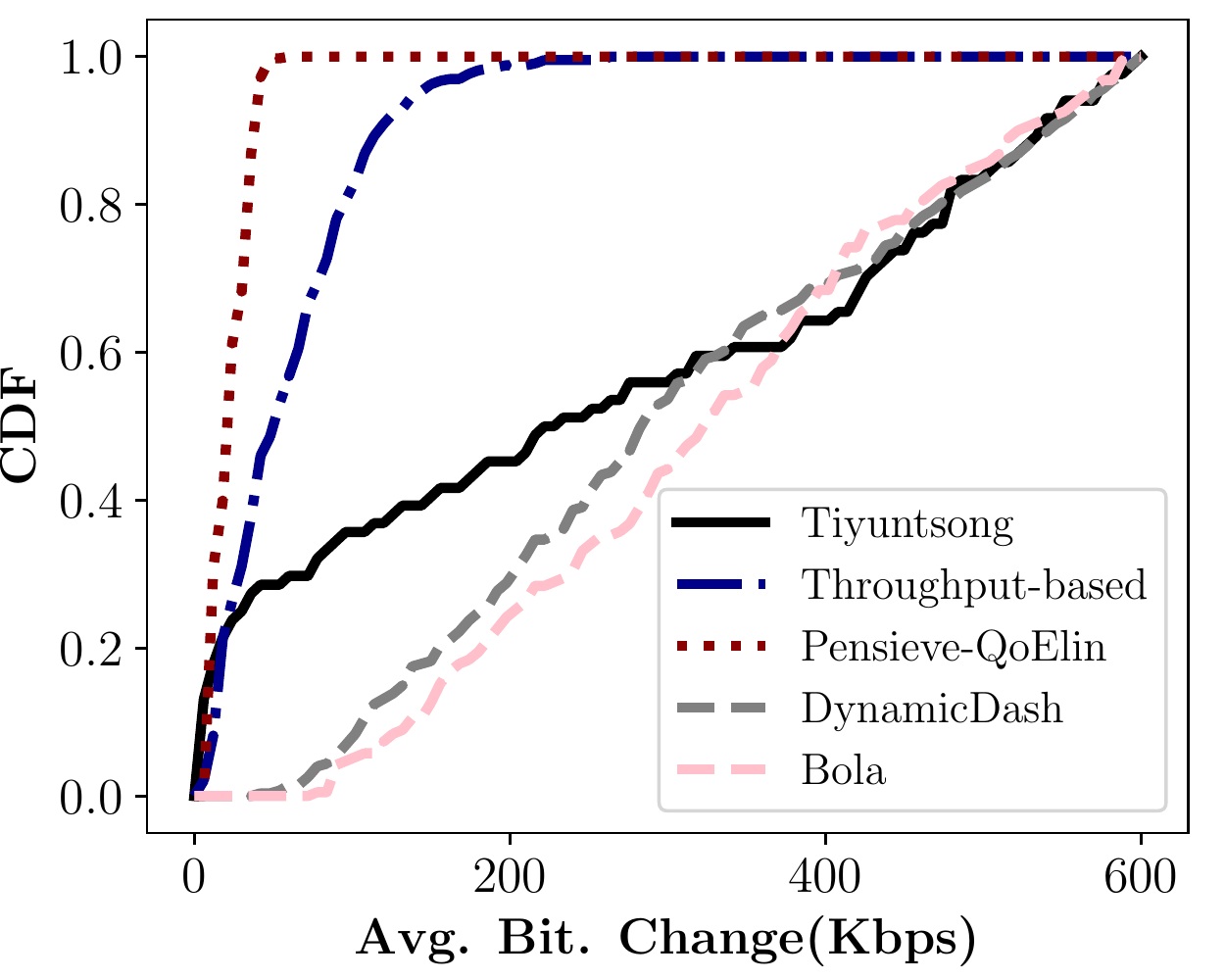}
    \caption{Comparing Tiyuntsong with existing ABR approaches on the same network traces. Results is shown with the win rate and the distribution of average bitrate, rebuffering time and average bitrate change for the approaches. Results show that Tiyuntsong wins existing approaches, with the win rate of 62\% to 100\%.}
    \label{fig:tiyuntsongvsabr} 
\end{figure}

\subsubsection{Tiyuntsong vs. Existing ABR Approaches}
\label{sec:tivsothers}
In this experiment, we aim to evaluate the Elo Ratings of several existing ABR algorithms including BOLA, DynamicDASH, Throughput-based, Constrained, and Pensieve~(QoE-lin). BOLA and DynamicDASH have been implemented in~\cite{spiteri2018theory}, and we use the harmonic mean of past five throughput measured to present the throughput-based rule. Moreover, we denote the constrained rule to select the intermediate chunk of the next video chunks and train a model via Pensieve optimized by \texttt{QoE-lin}~\cite{yin2015control}. We train Tiyuntsong about 2,000 epochs on the network traces datasets. We use 80\% dataset for training, 20\% for validating and leverage Oboe dataset for testing. Figure~\ref{fig:tytvs} shows the performance of training Tiyuntsong for 2,000 steps, we observe that Tiyuntsong performs better than the existing approaches after 1,800 steps. We also report Tiyuntsong's win rate and the CDF distribution of three underlying metrics in Figure~\ref{fig:tiyuntsongvsabr}. Compared to DynamicDash, Tiyuntsong improves the average bitrate by 3.19\%, decreases the average rebuffering time by 4.92\%, and reduces the 95th percentile average bitrate change by 16.47\% respectively. As expected, we also observe that Pensieve does reach an overwhelming advantage on the QoE metric but fails to perform well under some underlying metrics such as average bitrate, and it also proves our motivation of this work.


\vspace{-5pt}
\section{Conclusions and Future Work}
We propose Tiyuntsong, self-play RL approach to select bitrates for next video chunk. Unlike previously proposed approaches, Tiyuntsong uses two agents to compete against each other for automatically generating a better ABR algorithm. Experimental results prove that Tiyuntsong has achieved the state-of-the-art ABR algorithm via self-play. Additional research may focus not only to accelerate the training process but also to extend our work to solve the general incomplete information game problem. 

\section*{Acknowledgements}
We thank the anonymous reviewers for their valuable feedback. I also thank my wife Yuyan Chen for her great supports and brave ideas.
The work is supported by the National Key R\&D Program of China~(2018YFB1003703)，NSFC under Grant 61521002, Beijing Key Lab of Networked Multimedia, Alibaba-Tsinghua Joint Project~(20172001689).

\begin{small}
\bibliographystyle{IEEEbib}
\bibliography{main.bbl}
\end{small}
\begin{appendices}
\section{Appendix}
This supplementary material details the principle of Tiyuntsong\footnote{In memory of Jin Yong~(1924 - 2018).}. Due to the length of the supplemental material, we list a content to facilitate the selection of interested parts for review. Although these contents have \textbf{NOT} appeared in the main text, we believe that they will help the reviewer get a thorough understanding of Tiyuntsong.
\begin{itemize}
    \item Related work including adversarial learning~(\S\ref{sec:al}) and ABR's background~(\S\ref{sec:background});
    \item Tiyuntsong's network architecture~(\S\ref{sec:Details}), training procedure~(\S\ref{sec:Procedure}); 
    \item Tiyuntsong's training time~(\S\ref{sec:time}), another experiment and result for evaluating Tiyuntsong's performance under different network architecture candidates~(\S\ref{sec:Experiments});
    \item Additional discussions: The relationship between traditional QoE functions and Rules~(\S\ref{sec:discussions}), Why these network traces are selected?~(\S\ref{sec:diversity})
\end{itemize}
\section{Related Work}
\subsection{Adversarial Learning}
\label{sec:al}
Since GAN first proposed~\cite{goodfellow2014generative}, the adversarial discriminative learning method has been widely used in the various fields.
~The original GAN model is short of the loss function. Thus, many approaches, such as LSGAN~\cite{mao2017least} and WGAN-gp~\cite{gulrajani2017improved}, extend GAN's training methodology with strong theoretical proof. 
Moreover, adversarial learning has also been applied to extract features. For example, CoGAN~\cite{NIPS2016_6544} uses GAN to solve the domain transfer issue, and ADDA~\cite{tzeng2017adversarial} proposes a GAN-based generalized framework for domain adaptation.

\subsection{Background on ABR}
\label{sec:background}
\begin{figure}
    \centering
        \centering
          \includegraphics[width=0.8\linewidth]{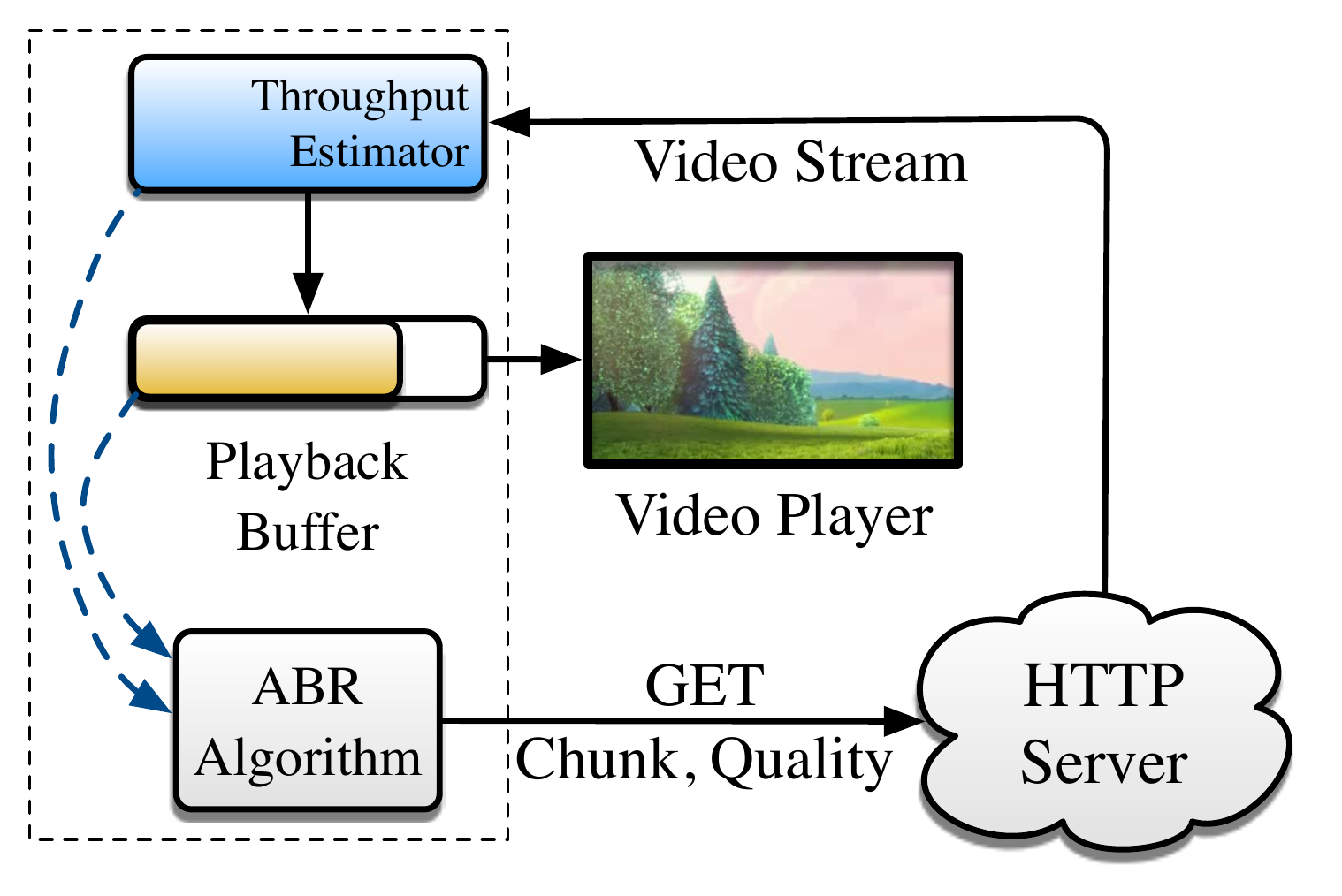}
        \caption{An overview of ABR video streaming.}
        \label{fig:abr}
\end{figure}

Due to the rapid development of network services, watching video streaming online has become an upcoming trend. Today, adaptive video streaming, such as HLS~(HTTP Live Streaming) and DASH, an algorithm that dynamically selects video bitrates via network conditions and client's buffer occupancy, is the predominant form of video delivery. The traditional video streaming architecture is shown in Figure~\ref{fig:abr}, which consists of a video player client with a constrained buffer length and an HTTP-Server or Content Delivery Network~(CDN). The video player client decodes and renders video frames from the playback buffer. Once the streaming service starts, the client fetches the video chunk from the HTTP Server or CDN orderly by an ABR algorithm, and, in the meanwhile, the ABR algorithm, implemented on the client side, determines the next chunk $N$ and next chunk video quality $Q_N$ via throughput estimation and current buffer utilization.
After finished to play the video, several metrics, such as total bitrate $b$, total re-buffering time $r$ and total bitrate change $s$ will be summarized as a QoE metric to evaluate the performance. Thus, achieving a high QoE score for video streaming has become a major challenge for ABR algorithms.

To overcome this challenge, most traditional ABR algorithms leverage time-series prediction~(throughput-based) or automation control method~(buffer-based) to make decisions for the next chunk. Moreover, \cite{mao2017neural} suggests that traditional fixed control rules methods require careful tuning and will achieve bad performances in the circumstance which is different from the assumption. As a result, traditional ABR algorithms perform well in pre-assumption and specific network conditions but hard to keep its performance in various network environments.

\section{Tiyuntsong's Details}

\subsection{Architecture and Implementation Details}
\label{sec:Details}
\begin{figure}
    \centering
      \includegraphics[width=0.9\linewidth]{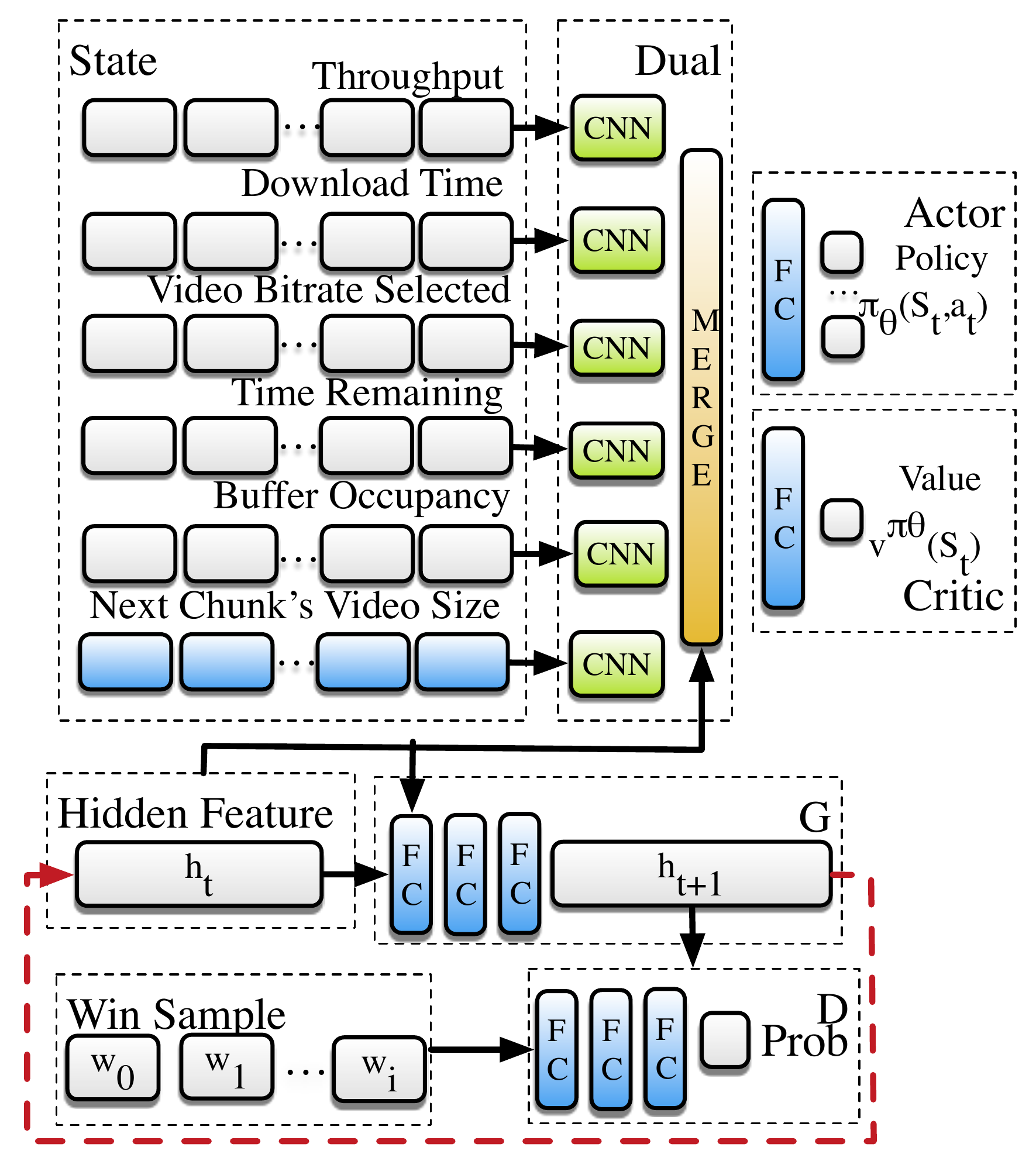}
    \caption{Tiyuntsong's network architecture.}
    \label{fig:network} 
\end{figure}

We use TensorFlow~\cite{abadi2016tensorflow} to implement Tiyuntsong\footnote{[Online]Available: \texttt{https://github.com/\textbf{anonymous}}}. As demonstrated in Figure~\ref{fig:network}, Tiyuntsong is composed of five neural network architectures as follows.

\textbf{Dual network:} We set past sequence length $k=10$. Features are extracted from the input state via a feature extraction layer. For each feature in the input state, it's passed through a conv-1d layer with 64 filters and the kernel size of $1\times3$. Meanwhile, we use \emph{ReLU} function as the activation function after each layer. Finally, the feature maps are concatenated as a tensor. 

\textbf{Policy network \& Value network:} Both policy network and value network are performed behind the Dual network. We use a fully connected layer with 64 neurons and active function \emph{ReLU} to represent them. The output of each network is n-dim vector and a single scalar respectively. In this work, we set $\gamma=0.6$, $\beta=0.01$, the learning rate for policy network $\alpha_0=10^{-4}$, and the learning rate for value network $\alpha_v=10^{-3}$. In this experiment, we use Adam optimizer~\cite{kingma2014adam} with default parameters to optimize these neural networks.

\textbf{Generative network \& Discriminator network:} Like previous work, the generative network and discriminator are composed of fully connected layer $\texttt{FC}$ and batch normalization layer $\texttt{BN}$. The generative network architecture is described as $\texttt{FC}_{64}^{1} \rightarrow \texttt{BN}^{1} \rightarrow \texttt{FC}_{32}^{2} \rightarrow \texttt{BN}^{2} \rightarrow \texttt{FC}_{16}^{3}$, and the discriminator network is listed as $\texttt{FC}_{64}^{1} \rightarrow \texttt{BN}^{1} \rightarrow \texttt{FC}_{32}^{2} \rightarrow \texttt{BN}^{2} \rightarrow \texttt{FC}_{1}^{3}$. Meanwhile, we use \emph{Leaky ReLU} as the active function and set learning rate for the generative network and discriminator network $\alpha_G=\alpha_D=10^{-4}$, hidden feature size ${size}_{h_{t}}=16$. Referring to the recommendations in LSGAN~\cite{mao2017least}, we use RMSProp optimizer~\cite{tieleman2012lecture} to update their gradients.

\subsection{Tiyuntsong's Training Procedure}
\label{sec:Procedure}
See details in Algorithm~\ref{alg:Overall}.
\begin{algorithm}
\caption{Tiyuntsong's Overall Training Procedure} 
\label{alg:Overall} 
\begin{algorithmic}[1]
\Require The ABR environment $\texttt{Env}$ to measure total bitrate, total rebuffer time, and total bitrate change with given video samples $\mathbf{V}$ and network traces $\mathbf{T}$; Two agents with GAN Enhance Module $\{\mathbf{A}_0;\mathbf{G}_0,\mathbf{D}_0\}$ and $\{\mathbf{A}_1;\mathbf{G}_1,\mathbf{D}_1\}$; A rule for estimating reward $\texttt{Rule}$;.
\Procedure{Training}{$\texttt{Env}, \mathbf{T}, \mathbf{V}, \mathbf{A}_0, \mathbf{A}_1$}
\State Initialize the parameters $\theta$ of $\theta_{\mathbf{A}_0}$ and $\theta_{\mathbf{A}_1}$ with random weights respectively.
\Repeat
\Comment{$\texttt{Epoch} \gets \texttt{Epoch} + 1$}
\State Sample $\texttt{trace}, \texttt{video}$ from $(\mathbf{T}, \mathbf{V})$;
\For{$(t, v) \in (\texttt{trace}, \texttt{video})$}
\State $(s_0, a_0) \gets \texttt{Env}(\mathbf{A}_0,t,v);$
\State $(s_1, a_1) \gets \texttt{Env}(\mathbf{A}_1,t,v);$
\EndFor
\State Compute reward $\mathbf{R}\in\{r_0,r_1\}\gets\texttt{Rule}(s_i, a_i)$;
\State Estimate winning percentage $\mathbf{w} \in \{w_0, w_1\}$ from $\mathbf{R}$;
\For{$i \in \{0,1\}$}
\State Get winning samples $\mathbf{A}_{i_w}$;
\State Update $\mathbf{D}_i$ and $\mathbf{G}_i$ with (\ref{eq:dloss}) and (\ref{eq:gloss}) using ~($s_i,a_i,r_i,\mathbf{A}_{i_w}$);
\State Update policy with (\ref{eq:1}) and (\ref{eq:2}) using ($s_i,a_i,r_i,w_i$);
\EndFor
\Until{Converged}
\EndProcedure
\end{algorithmic} 
\end{algorithm}




\subsection{Tiyuntsong meets Parallel Training}
\label{sec:parallel}
During the training process, we observe that the training progress is inefficient while using a single process. Inspired by the multi-agent training method~\cite{mnih2016asynchronous}, we modify Tiyuntsong's training in the single agent as training in multi-agents. Multi-agents training consists of two parts, a central agent and a group of forwarding propagation agents. The forward propagation agents only decide with both policy and critic via state inputs and neural network model received by the central agent for each step; then it sends the $n$-dim vector containing $\{state, action, reward, gan\}$ to the central agent. The central agent uses the actor-critic algorithm to compute gradient and then updates its neural network model. Finally, the central agent pushes the updated network parameters to each forward propagation agent. Note that this can happen asynchronously among all agents, for instance, there is no locking between agents. By default, Tiyuntsong with multiple training uses 12 forward propagation agents and one central agent;

\section{Additional Evaluations}
Instead of original paper, We still evaluate Tiyuntsong under various neural network architectures. Our results answer the questions: How long will Tiyuntsong converge? What's the best neural network architecture for Tiyuntsong?

\subsection{Tiyuntsong's Training Time}
\label{sec:time}
Tiyuntsong trains itself via endless competition, so the longer Tiyuntsong trains, the better it performs. In this paper, we stop training Tiyuntsong on i7-4790k CPU in 4 cores till its Elo-rating exceeds previous approaches. (Unlike traditional CV work, AI in networking requires a small model which can obtain high performance in low costs, so training on CPU is feasible). The training time lasts about 1.5 days. We observe that Tiyuntsong outperforms previously proposed approaches in 40mins, 2hrs, 13hrs, and 33hrs respectively. We will focus on accelerating the training process as is described in future work.

\renewcommand\arraystretch{1.0}

\begin{table}
    \centering
        \begin{tabular}{ccc}
            \toprule
            \textbf{Architecture} & \textbf{Elo} & \textbf{Timespan(it/s)}\\
            \midrule
            FC & 1033 & \textbf{1.28} \\
            LSTM & 1057 &  0.77 \\
            2D-CNN & 1040 & 1.16 \\
            1D-CNN & \textbf{1094} & 1.04 \\
            \midrule
            Constrained & 977 & - \\
            Throughput-Rule & 1023 & - \\
            \bottomrule
        \end{tabular}
        \caption{Comparing performance (Elo ratings) of Tiyuntsong with different neural network architectures including Fully Connected, 2D-CNN, 1D-CNN and LSTM. Results are evaluated under same network traces and video description in 50 steps.}
        \label{tbl:tytarch}
\end{table}
\begin{figure}[ht]
    \centering
      \includegraphics[width=0.9\linewidth]{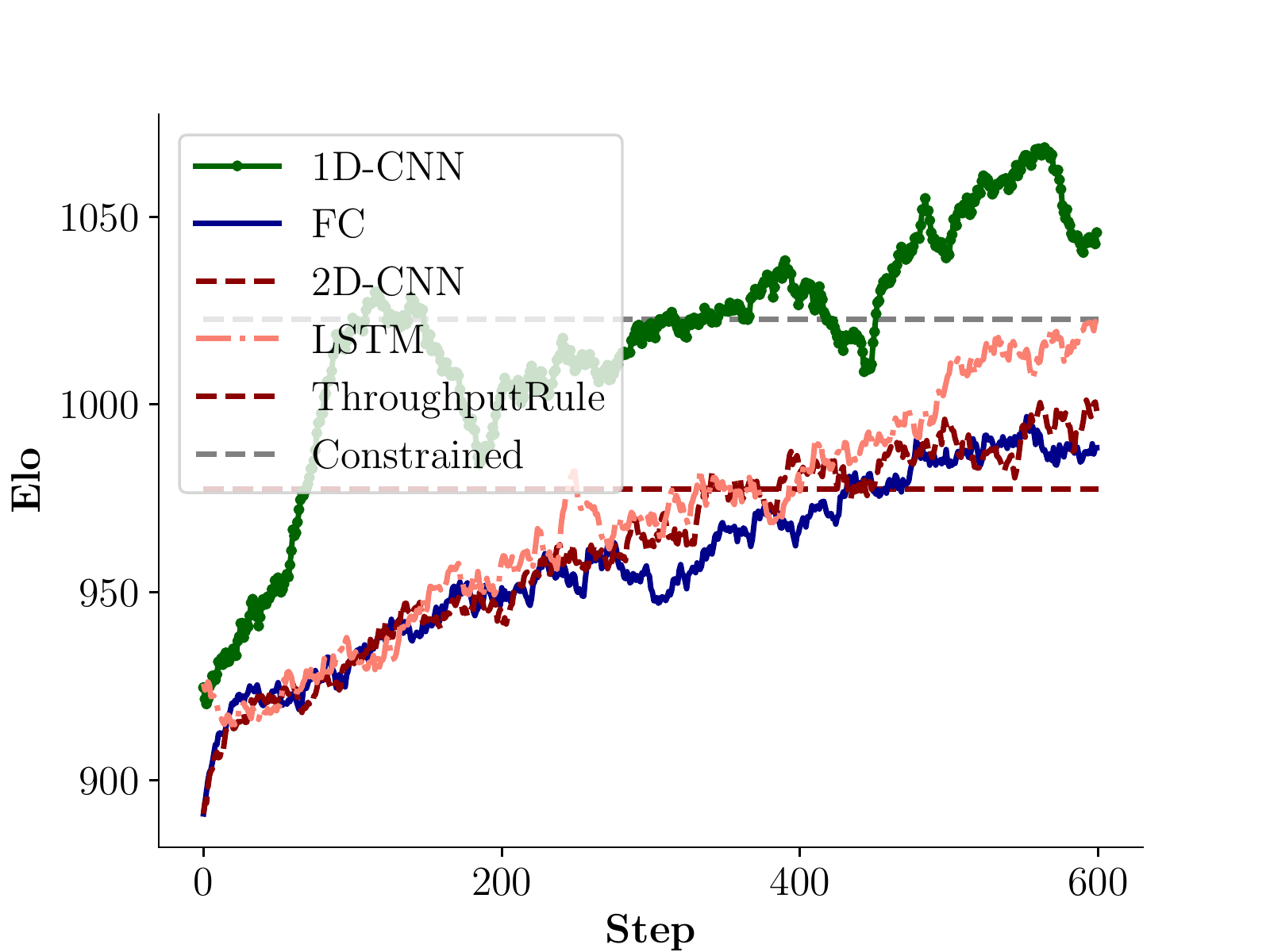}
    \caption{Tiyuntsong's network architecture.}
\end{figure}
\subsection{Tiyuntsong with Different Architectures}
\label{sec:Experiments}
In this experiment, we compare the Dual network architecture from Tiyuntsong to the following network architectures which collectively represent the architecture candidates. The network architecture candidates are simply listed as follows:

\begin{itemize}
    \item \textbf{Fully Connected} 
    
    $\texttt{FC}_{64}^{1} \rightarrow \texttt{FC}_{128}^{2} \rightarrow \texttt{FC}_{64}^{3}$ 
    
    \item \textbf{LSTM~(long-short-term-memory)} 
    
    $\texttt{LSTM}_{64}^{1} \rightarrow \texttt{LSTM}_{64}^{2} \rightarrow \texttt{SELF-ATTENTION}_{64}^{1}$
    
    \item \textbf{2D-CNN} 
    
    $\texttt{CONV2D}_{64}^{1} \rightarrow \texttt{MAXPOOL}_{2}^{1} \rightarrow \texttt{CONV2D}_{64}^{2} \rightarrow \texttt{MAXPOOL}_{2}^{2} \rightarrow \texttt{FC}_{64}^{1}$
    
    \item \textbf{1D-CNN$^{*}$}
    
    $\texttt{CONV1D}_{64}^{1\cdots6} \rightarrow \texttt{MERGE}^{1} \rightarrow \texttt{FC}_{64}^{1}$
\end{itemize}

We train and test under Sabre environment with the same network traces and video descriptions. In this experiment, we set $\gamma=0.99$, $\beta=0.02$, $step=50$ for only testing their performance instead of convergence. We report the result in Figure~\ref{fig:tytarch} and Table~\ref{tbl:tytarch}, where 1D-CNN is the Tiyuntsong's Dual network architecture. The obtained results indicate that 1D-CNN neural network architecture succeeds in improving the Elo ratings, with improvements in average Elo ratings of 37 - 61. We also observe that there is no obvious difference between these architectures in terms of operational efficiency.

\section{Discussions}
\begin{figure}
    \centering
    \includegraphics[width=0.4\linewidth]{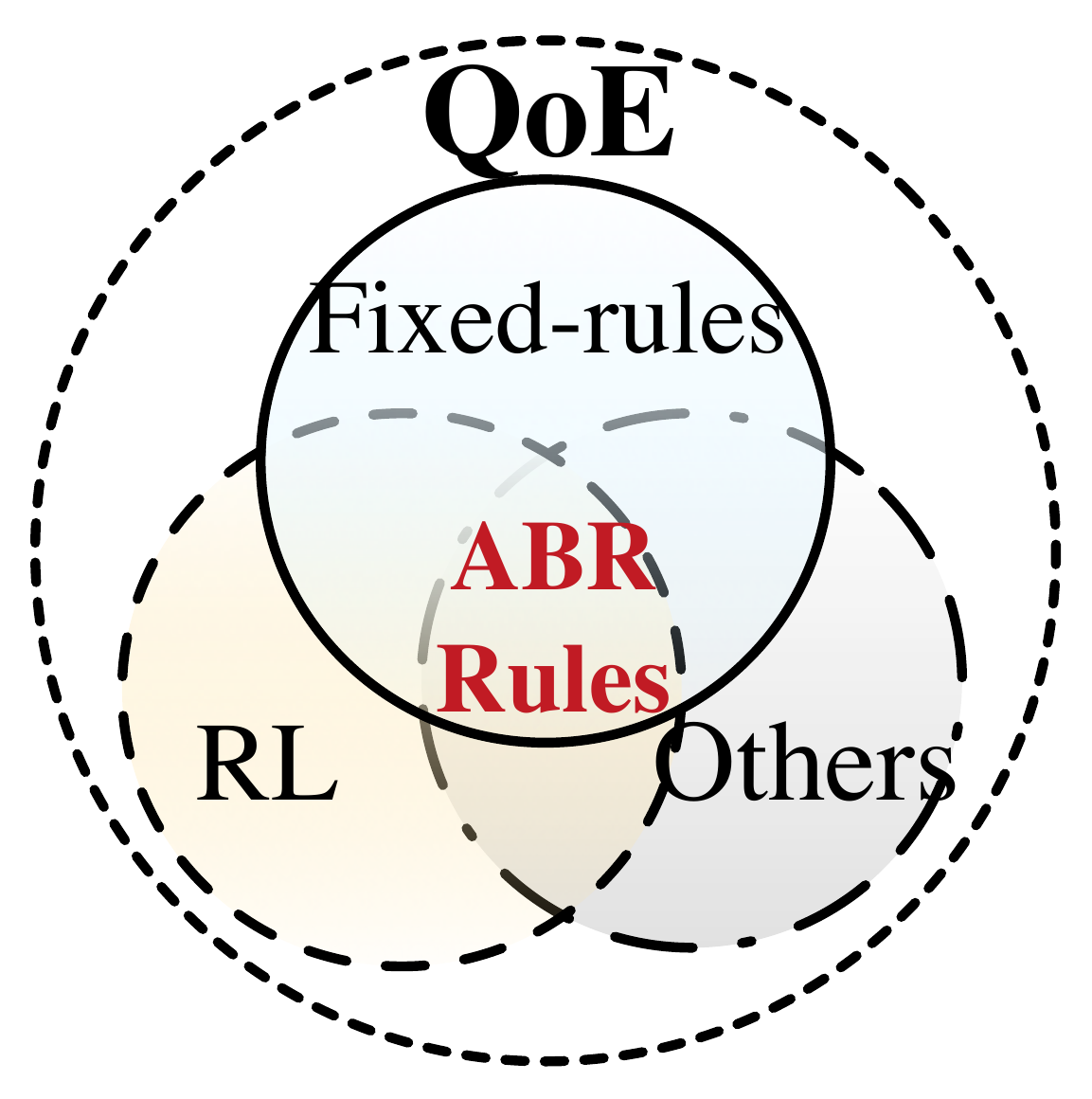}
    \caption{Traditional RL method's Trap: A QoE metric can evaluate several ABR algorithms, but the generated algorithm may deviate from the basic rules of the ABR if it blindly improves QoE score.}
    \label{fig:qoe} 
\end{figure}
\subsection{Traditional QoE functions and Rules}
\label{sec:discussions}
As demonstrated in Figure~\ref{fig:qoe}, we find that more than one rule can be generalized to represent the same QoE formulation, vice versa. For example, during the design of method, fixed-rules, such as throughput-based and buffer-based, use handcraft features or network presumptions to implement the model without considering how to take advantage of evaluation metrics~(QoE formulation). Then, the given QoE formulation is only used to evaluate the performance of each algorithm. Furthermore, mixed-based and RL-based schemes, i.e., MPC~\cite{yin2015control} and Pensieve~\cite{mao2017neural} adopts the QoE formulation to \emph{guide} its algorithm for achieving higher QoE score. However, recent research exposes that there still exists a plenty of room for improving QoE metrics and many situations (e.g., some network conditions and videos) cannot be evaluated correctly via current QoE metrics due to the lack of features, as the RL-based scheme still tries to optimize the QoE score with the false guidance, finally results in failure of real-world performances. As a result, no matter how precisely and carefully the QoE function tunes, traditional RL-methods cannot exactly provide a result that the users desire. 

Intuitively, the critical idea of Rule is to tackle the problem that the reward function fails to depict. For example, ABR tasks and self-driving car tasks. The fundamental factor of Rule is: Given two answers~(action) from one questions~(state), can \emph{you} figure out which one is better to answer? In this paper, we prove that using self-play reinforcement learning will learn the strategy by itself if \emph{you} can \emph{tell} the agent who is better.

\subsection{The Diversity of Network Traces}
\label{sec:diversity}
The real world network is composed of several network conditions such as 3G/HSPDA, 4G, Wired and WiFi. It's obvious that each of them has different features, and we try to train a generalized model which can cover all the network status. Thus, we collect a corpus of network traces by combining several public datasets. Meanwhile, the diversity of the length of network traces is still challenging. On the one hand, each dataset is generated in different durations and granularity. For example, each of FCC dataset we used logs the average throughput about 100 seconds, and at a 5-second granularity (the log is sized 20); Each of synthetic network traces is logged as the average throughput about 2000 seconds. On the other hand, we balance the data distribution by controlling the amount of various network traces in the data pool during the training process. Additional information will be open-sourced later on. 
\end{appendices}
\end{document}